\documentclass[sigconf]{acmart}
\usepackage{algorithm}
\usepackage{verbatim}
\usepackage{hyperref}
\usepackage{algpseudocode}

\usepackage{nomencl}
\makenomenclature

\setcopyright{none}
\renewcommand\footnotetextcopyrightpermission[1]{} 
\AtBeginDocument{%
  \providecommand\BibTeX{{%
    \normalfont B\kern-0.5em{\scshape i\kern-0.25em b}\kern-0.8em\TeX}}}

\begin{document}

\title{Quantum-Enhanced Selection Operators for Evolutionary Algorithms}

\author{David Von Dollen}
\authornote{Corresponding author email: David.VonDollen@vw.com}
%\authornote{Also affiliated with Leiden University}

%\email{webmaster@marysville-ohio.com}
\affiliation{
  \institution{Volkswagen Group of America}
%  \streetaddress{P.O. Box 1212}
  \city{Auburn Hills}
  \country{USA}
}

%\author{Lars Th{\o}rv{\"a}ld}
%\affiliation{%
%  \institution{The Th{\o}rv{\"a}ld Group}
%  \streetaddress{1 Th{\o}rv{\"a}ld Circle}
%\email{larst@affiliation.org}

\author{Sheir Yarkoni}
%\authornote{Also affiliated with Leiden University}
\affiliation{%
  \institution{Volkswagen Group}
  \city{Munich}
  \country{Germany}
}

\author{Daniel Weimer}
\affiliation{%
  \institution{Volkswagen Group}
  \city{Munich}
  \country{Germany}
}

\author{Florian Neukart}
%\authornote{Also affiliated with Leiden University}
\affiliation{
  \institution{Terra Quantum}
  \city{Zurich}
  \country{Switzerland}
}

\author{Thomas Bäck}
\affiliation{%
  \institution{Leiden University}
  \city{Leiden}
  \country{Netherlands}
}

%\affiliation{%
% \institution{Rajiv Gandhi University}

%\affiliation{%
%  \institution{Tsinghua University}
%  \streetaddress{30 Shuangqing Rd}

%\affiliation{%
%  \institution{Palmer Research Laboratories}
%  \streetaddress{8600 Datapoint Drive}
%  \country{USA}

%\affiliation{%
%  \institution{The Th{\o}rv{\"a}ld Group}
%  \streetaddress{1 Th{\o}rv{\"a}ld Circle}
%\email{jsmith@affiliation.org}

%\affiliation{%
%  \institution{The Kumquat Consortium}
%  \city{New York}
%\email{jpkumquat@consortium.net}

\renewcommand{\shortauthors}{Von Dollen, et. al}

\begin{abstract}
Genetic algorithms have unique properties which are useful when applied to black box optimization. Using selection, crossover, and mutation operators, candidate solutions may be obtained without the need to calculate a gradient.  In this work, we study results obtained from using quantum-enhanced operators within the selection mechanism of a genetic algorithm. 
Our approach frames the selection process as a minimization of a binary quadratic model with which we encode fitness and distance between members of a population, and we leverage a quantum annealing system to sample low energy solutions for the selection mechanism.  We benchmark these quantum-enhanced algorithms against classical algorithms over various black-box objective functions, including the OneMax function, and functions from the IOHProfiler library for black-box optimization. We observe a performance gain in average number of generations to convergence for the quantum-enhanced elitist selection operator in comparison to classical on the OneMax function.  We also find that the quantum-enhanced selection operator with non-elitist selection outperform benchmarks on functions with fitness perturbation from the IOHProfiler library. Additionally, we find that in the case of elitist selection, the quantum-enhanced operators outperform classical benchmarks on functions with varying degrees of dummy variables and neutrality. 
\end{abstract}




%  \includegraphics[width=\textwidth]{teaser}
%  \caption{Cartoon of encoding and workflow for Quantum-Enhanced Selection Operator}
%  \Description{Cartoon of encoding and workflow for Quantum-Enhanced Selection Operator}

\begin{teaserfigure}
  \includegraphics[width=\textwidth]{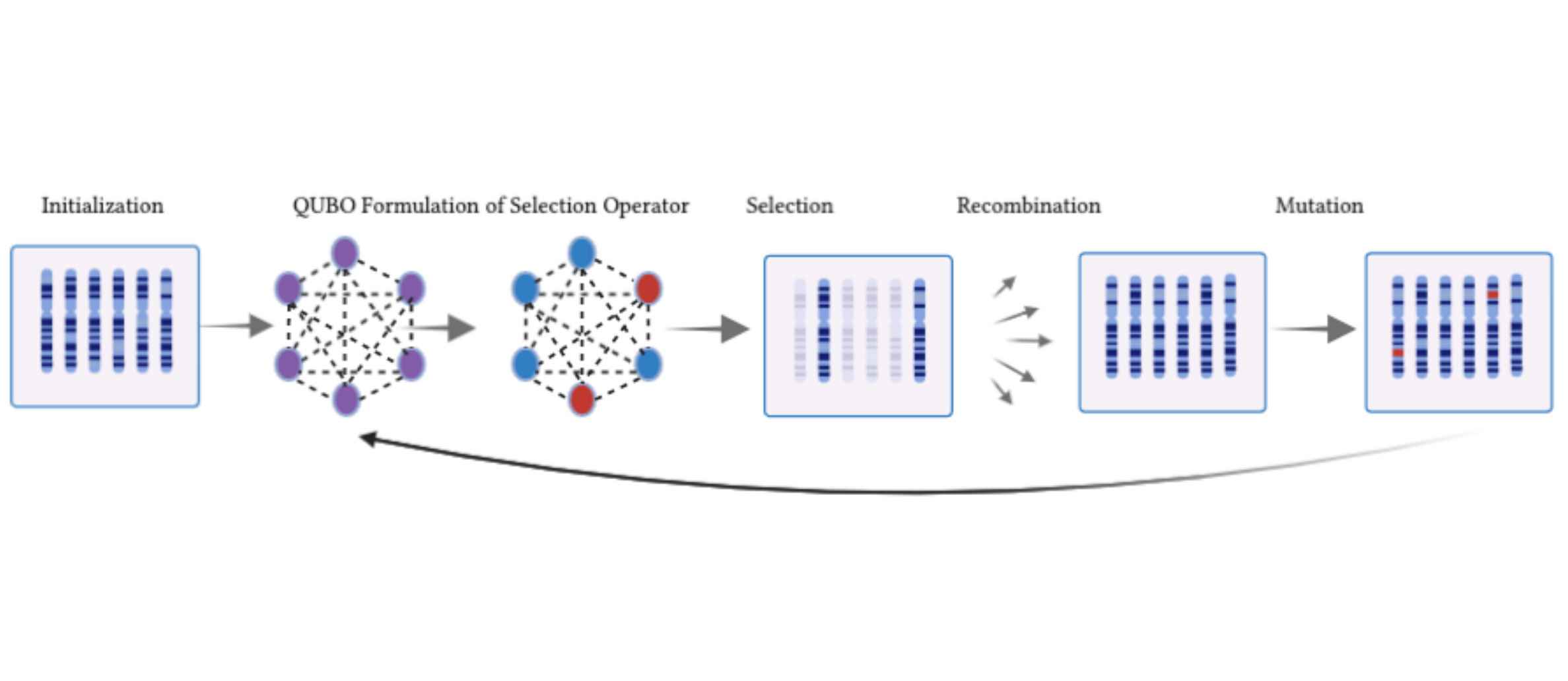}
  \caption{Schematic of encoding and workflow for Quantum-Enhanced Selection Operator for a Quantum-Enhanced Evolutionary Algorithm}
  \Description{Schematic of encoding and workflow for Quantum-Enhanced Selection Operator for a Quantum-Enhanced Evolutionary Algorithm}
  \label{fig:teaser}
\end{teaserfigure}

\maketitle

\section{Introduction}
%%ACM's consolidated article template, introduced in 2017, provides a
%consistent \LaTeX\ style for use across ACM publications, and
%incorporates accessibility and metadata-extraction functionality
%necessary for future Digital Library endeavors. Numerous ACM and
%SIG-specific \LaTeX\ templates have been examined, and their unique
%features incorporated into this single new template.

%If you are new to publishing with ACM, this document is a valuable
%guide to the process of preparing your work for publication. If you
%have published with ACM before, this document provides insight and
%instruction into more recent changes to the article template.

%The ``\verb|acmart|'' document class can be used to prepare articles
%for any ACM publication --- conference or journal, and for any stage
%of publication, from review to final ``camera-ready'' copy, to the
%author's own version, with {\itshape very} few changes to the source.
%Discuss evolutionary computation.
Evolutionary algorithms are nature-inspired models drawn for observations of organic evolution. Using selection, recombination, and mutation, this family of algorithms evolves populations to search an optimization domain with respect to the individual fitness within the population. Related to the fields of biology, numerical optimization, and artificial intelligence, these algorithms may also model a collective learning process, where individuals may not only represent a point on the domain of the objective function, but also may represent knowledge of an environment \cite{Bck1996EvolutionaryAI}. This family of algorithms are particularly well suited for Black-box optimization, where there is no knowledge of the internal structure of the objective function, in that there is no need to calculate a gradient to search the landscape, which could be highly nonlinear and rugged with many local optima.

Quantum computing has steadily shown an increasing potential for disrupting limitations imposed by seemingly intractable problems \cite{Kanamori}. By leveraging properties of quantum mechanics, such as superposition, entanglement and interference, researchers realized theoretical speedups for processes such as order finding, factoring, and search \cite{simons, Shor1999PolynomialTimeAF, grovers}. In our current era of noisy-intermediate scale quantum devices (NISQ) \cite{NISQ}, as quantum computing hardware continues to evolve, it is not without growing pains as it has been realized that coherence times and fault tolerance to external noise are not perfect. For algorithms with a provably exponential speedup over classical algorithms, hardware with many physical qubits (quantum bits), low noise, and long coherence times to may be required to realize circuits with the required depth and scale. However, in the near term, there exists the possibility to investigate if there are sub-components of routines for classical algorithms that may be intractable for classical computation which may benefit from leveraging qualitative performance enhancements from a NISQ-era quantum computing systems leveraging quantum effects.

In this work we examine leveraging a quantum annealing system to find solutions to the problem of optimal selection within an evolutionary algorithm, encoded as a binary quadratic model. We examine the trade-off in selective pressure vs. exploration in the evolutionary search, and show qualitative gains with respect to fitness and expected run-times in the form of average generations to convergence. We investigate these performance gains with respect to  the change in the ratio of $\mu$ to $\lambda$, or the size of the selected parent pool and number of offspring, and find that the gap in performance grows as the ratio approaches $\lambda$/$\mu$=2. 
%\begin{figure*}
%  \includegraphics[width=\textwidth]{teaser}
%  \caption{Schematic of encoding and workflow for %Quantum-Enhanced Selection Operator for a Quantum-Enhanced %Evolutionary Algorithm}
%  \Description{Schematic of encoding and workflow for %Quantum-Enhanced Selection Operator for a Quantum-Enhanced %Evolutionary Algorithm}
%\end{figure*}
This is not without a cost, as we also observe an additional overhead in compute times, which are incurred by making calls across a network to query a quantum processing unit at each generation, an issue also identified in \cite{sharabiani2021quantum}. We also confirm findings the authors reported in this work, where for fully connected graphs of input QUBOs (Quadratic Unconstrained Binary Optimization) constructed from randomly initialized populations, hybrid quantum-classical outperform fully quantum solvers by finding lower energy configurations given the input. %However if we discount these and consider only the anneal times, which are constant, and the time to construct the input for the binary quadratic model, then the overhead in time complexity is at most polynomial with respect to the input.

We test our quantum-enhanced algorithms on the IOHprofiler suite for black-box optimization, specifically examining Pseudo-Boolean functions \cite{IOHprofiler}.We find that for functions with perturbed fitness, quantum-enhanced selection operators achieve slightly better performance to their classical counterparts. We find that our quantum enhanced algorithms generally match or outperform their classical counterparts on a majority of test functions, 10 out of 15 test functions, tested for significance with p-values < 0.05.
\section{Related Works}
%Discuss similar works in quantum-inspired optimization
Quantum-inspired evolutionary algorithms have been well studied over the years, starting with \cite{542334}, where classical simulation of quantum mechanical properties were applied to evolutionary search. This culminated in a large body of work with many variants of quantum-inspired algorithms as described by Zhang in \cite{Zhang2011}.  

In surveying quantum-inspired algorithms, Zhang noted 3 types of algorithms which combine quantum computational properties with evolutionary algorithms \cite{Zhang2011}. These include:
 \begin{itemize}

\item Evolutionary Designed Quantum Algorithms (EDQA), which leverage evolutionary algorithms to evolve new designs of quantum algorithms

\item Quantum Evolutionary Algorithms (QEA), where evolutionary algorithms are implemented on a quantum computer

\item Quantum-Inspired Evolutionary Algorithms (QIEA), which are algorithms where the evolutionary process is supplemented by routines inspired by quantum mechanics, but implemented using classical hardware.

\end{itemize}
Along with these, we propose to consider an additional algorithm class. As we are currently in the NISQ era for quantum hardware, we can also examine hybrid quantum-classical algorithms, where some portions or subroutines of the algorithm's execution are performed on a quantum computer, and other portions are performed classically. We call this type Quantum-Enhanced Evolutionary Algorithms, and give an example illustration of this concept in Fig. 1.

Studies into quantum-enhanced evolutionary algorithms can be examined from a standpoint of leveraging a quantum device and quantum mechanical properties for selection, crossover, or mutation operators within the heuristic of the evolutionary search. 

Of these evolutionary operators, the idea for a genetic algorithm assisted by quantum annealing was proposed by Chancellor in \cite{ChancellorGenetic}, and the authors of \cite{king2019quantumassisted} investigated using a quantum-assisted mutation operator, and leveraging reverse annealing runs using a quantum annealer. By performing qausi-local searches using the quantum-assisted mutation operator, the authors were able to show an improvement over forward quantum annealing in finding global optima for a set of input spin-glasses. More recently, investigation into continuous black box optimization leveraging a quantum-assisted acquisition function have been reported in \cite{izawa2021continuous}. In \cite{sharabiani2021quantum}, the authors leverage a quantum annealing system to formulate continuous optimization problems cast within a quantum nonlinear programming framework, and show applications within the green energy space. Sharabiani et. al  also identified the overhead in compute times in regards to querying a QPU for an subroutine for their optimization algorithm in their work. To our knowledge, there has be no prior investigation into quantum-enhanced routines for selection, which our work addresses.

Turning our attention away from quantum annealing for a moment, there also exists a stream of research into applying Grovers's algorithm for unstructured search to global optimization \cite{Baritompa}. While we may not have systems of the scale and fidelity available today to implement these algorithms on a practical level, this stream of research could be further investigated and realized as quantum systems come online with higher orders of available error corrected qubits and longer coherence times.

 To motivate our work, we frame the selection process as a Maximum Diversity problem, where we wish to select a subset of parents for crossover and/or mutation, which preserve a high degree of quality diversity within the parent pool, with low genotype similarity between parents, while preserving a high degree of fitness in regards to the objective function. Maximum Diversity Problems have been shown to be NP-Hard \cite{Duarte07tabusearch, DeAndrade}. The difference in our formulation from classic Maximum Diversity Problems is that we do not use Euclidean distance as our distance function, but investigate other distance functions such as Hamming distance. This type of combinatorial optimization scales with respect to the input $n$, and is  dominated by $\mathcal{O}(2^n)$, as there is a binary decision variable for each individual within the population. However, we propose that by leveraging a quantum processing unit, or other hybrid quantum-enhanced or quantum-inspired methods, we may be able to sample approximate solutions of quality in comparison to other techniques and heuristics. This is the main motivation for leveraging a quantum computing system for this work.
 
 \section{Methods}
\subsection{Evolutionary Algorithms}

Evolutionary Algorithms model processes observed in nature including natural selection, reproduction and mutation. In regards to global optimization, these processes are leveraged to evolve individual solutions with respect to an objective or fitness function. In the case of a Genetic Algorithm for Pseudo-Boolean Optimization, A population of individuals $P(0)$ := [$\mathbf{a_0}$, ... $\mathbf{a_n}$] is initialized, where the the values of $\mathbf{a} \in [0,1]$ are generated at random uniformly. At each generation, individuals (also referred to as chromosomes in the case of genetic algorithms) from a population are selected, recombined to produce offspring, and mutated, resulting in a new population. Over time individuals  converge to minima of the objective function to be optimized with respect to their fitness values, given by a black box objective function $\psi$. These properties make evolutionary algorithms powerful candidates for optimization for black box functions where the domain and modality of the function is unknown for both discrete, as in the case of pseudo-Boolean optimization, and continuous optimization.

\subsection{Selection Operators for Evolutionary Algorithms}
In selecting parents from a population pool for mutation and crossover, we may choose from a number of selection operators. These operators may be deterministic or probabilistic and can include:
\begin{itemize}
   % \item \textit{q-tournament selection} In $q$-tournament selection $q$ parents are selected with uniform probability over the population of candidates. From these $q$, the best parent is selected, and the process is repeated to create offspring via crossover and mutation. 
    
    \item \textit{($\mu - \lambda$) selection} In ($\mu - \lambda$) selection, $\mu$ individuals are selected based on the rank of their fitness values to create $\lambda$ offspring. In this case, only child offspring are included in the subsequent population.
    
    \item \textit{($\mu + \lambda$) selection} In ($\mu + \lambda$) selection, $\mu$ individuals are selected based on the rank of their fitness values to create $\lambda$ offspring. ($\mu + \lambda$) selection is elitist, meaning the parents are included in subsequent generations. In our experiments, we also tracked and recombined the best solution found so far with selected members of the population to create subsequent generations.
\end{itemize}
We choose these operators in comparison to our quantum-enhanced operator in order to compare and contrast the trade offs in exploration vs. exploitation in our evolutionary search.

\subsection{A Quantum-Enhanced Selection Operator}
Maximum diversity problems are characterized by selecting elements from a set which maximize diversity within the selected subset. Kuo et. al \cite{Kuo1993} gave a formulation for this set of problems as a binary quadratic model, which were also shown to be NP-Hard.

When given a set of candidates within a population pool, a natural question is how to select parents with a high degree of fitness, yet are also diverse from one another, with the idea that we want to be able to balance the trade off between exploitation and exploration in or selection mechanism. Low population diversity may lead to more localized search and premature convergence.  Similar to the feature subset selection problem in machine learning \cite{vondollen2021quantumassisted}, the problem of selecting the optimal subset of parents from a population pool can be framed as a binary quadratic model. In our formulation we start by defining a matrix $\mathbf{Q}$:

\begin{equation}
\mathbf{Q}_{ij} = 
    \begin{cases}
    - \alpha|\psi(\mathbf{a}_i)| & \text{if}\ i=j \\
     -\beta|dist(\mathbf{a}_i, \mathbf{a}_j)| & \text{if}\ i < j \ \\%and\ dist = hamming\\
    % \beta|dist(\mathbf{a}_i, \mathbf{a}_j)|, & \text{if}\ i < j\ and\ dist\ \neq hamming\\
    0,  \text{otherwise}
    
    \end{cases}
\end{equation}

Where $\psi(\mathbf{a}_i)$ is the fitness evaluated by the chromosome $\mathbf{a}_i$, and $dist(\mathbf{a}_i, \mathbf{a}_j)$ is the pair-wise distance metric between chromosomes within the population. Note that for this formulation the distance metric may be arbitrarily chosen by the practitioner, in our case we use \textit{hamming} distance, where we negate the quadratic term as bit strings with higher values for hamming distance may be more distant in the sampled hyper-cube. 

We introduce the terms $\alpha$ and $\beta$ as scaling constants, which we may use to adjust the optimization domain for the binary quadratic model for more or less ruggedness in order to leverage the effect of quantum tunneling. Depending on the choice of distance metric, one may want to use a negative value for the scaling term $\beta$ applied to the quadratic terms of the QUBO, in the case where a higher value for the distance metric represents a higher degree of similarity or correlation.
 We may tune the $\alpha$ parameter to increase or decrease the selective pressure, giving more or less weight to individuals with respect to their fitness. The $\beta$ term allows to increase or decrease the diversity in the population of selected individuals in regards to the distance between their chromosomes. Overall the action of the two terms help to maintain selective pressure while also achieving a balance of diversity within the selected population. 

The $\mathbf{Q}$ matrix acts as input for our resulting minimization problem, where we wish to find an optimal assignment of qubit values, represented as $\sigma_i$ where $\sigma  \in [0,1]$ and $i \in [1, ..., n]$ to indices of the population which is of size $n$, where we select members of the population with a value of $\sigma = 1$ of size $\mu$
\begin{equation}
E(\mathbf{\mathbf{\sigma}}) = \sum_{i\leq j}\mathbf{\mathbf{\sigma}}_i\mathbf{Q}_{ij}\mathbf{\sigma}_j  + (\sum_i\mathbf{\sigma}_i - \mu)^2
\end{equation}
Using this population subset of size $\mu$, we may then perform crossover and mutation classically, creating a new population and increment to the next generation. For our experiments we investigate elitist and non-elitist versions of the quantum-enhanced operator, where the parents are included in the former case and not in the latter case. We also include the heuristic of recombining the selected population with the best solution found so far in the elitist version for both classical and quantum-enhanced operators. We outline the pseudo-code for these algorithms in Fig. 2.

\begin{figure}
\begin{algorithm}[H]
	\caption{Genetic Algorithm with quantum-enhanced Selection Operator} 
	\begin{algorithmic}[1]
	\Require population size $n$
	\Require chromosome size $\nu$
	\Require maximum generations $t$
	\Require mutation probability $m$
	\Require Boolean flag $\textit{elitist}$
	\Require Boolean flag $\textit{quantum-enhanced}$

	%\State Initialize $best = [x_0, x_1,,, x_\nu]$, where $x$ $\in [0,1]$ and are drawn with uniform probability
	\State Initialize $P(0) := [\mathbf{a}_0, ..., \mathbf{a}_n]$, $\mathbf{a} = [x_0, x_1,,, x_\nu]$, where $x$ $\in [0,1]$ and are drawn with uniform probability, set $best = \mathbf{a}_0$
	\For{$\mathbf{a} \in P(0)$}
            \If{$\psi(\mathbf{a}) > \psi(best)$}
            \State $best = \mathbf{a} $
            \EndIf
    \EndFor
	\For{generation $g$=1, 2, ..., $t$}
	    \If {\textit{quantum-enhanced}}
	    \State Construct $\mathbf{Q}$ according to ( Eq. 1)
	    \State Sample solution vector $min(E(\mathbf{\mathbf{\sigma}}))$ according to ( Eq. 3)
	    \State Select $P_\mu(g)$ = $P_\sigma(g)$ where $\mathbf{\sigma}_n = 1$ for [$\sigma_1,..,\sigma_n$] in $\mathbf{\sigma}$
	    \Else:
    	    \If {$g=1$}
    	    \State Select $\mu$ parents from $P(0) : P_\mu(g)$
    	    \Else
    	    \State Select $\mu$ parents from $P(g) : P_\mu(g)$
    	    \EndIf
	    \EndIf
	    \If{\textit{elitist}}
	    \State Perform crossover on $best$ with $P_\mu(g)$ to generate $P_\lambda(g)$ offspring, add $P_\mu(g)$ to $P_\lambda(g)$
        \Else
        \State Perform crossover on $P_\mu(g)$ to generate $P_\lambda(g)$ offspring
        \EndIf

        \State Mutate $P_\lambda(g)$ according to probability $m$
        %\State $P(g) = P_\mu(g) \oplus P_\lambda(g)$
        \State $P(g) = P_\lambda(g)$
        \For{$\mathbf{a} \in P(g)$}
            \If{$\psi(\mathbf{a}) > \psi(best)$}
            \State $best = \mathbf{a} $
            \EndIf
        \EndFor
	\EndFor \\
	\Return{$best$}
	\end{algorithmic} 
\end{algorithm}
\caption{Pseudocode of quantum-enhanced genetic algorithm}
\end{figure}

\section{Experiments}
\subsection{Benchmarking Quantum, Hybrid, and Classical Solvers}
For our version of a quantum-enhanced evolutionary algorithm, we make calls to the D-Wave quantum annealer, using the quantum processing unit (QPU). During the annealing regime, the system starts in a state of superposition for all qubit values, and by gradually reducing the amplitude of a transverse field, drives the system to a ground state. By leveraging quantum-mechanical properties such as entanglement and superposition, we may observe an effect know as \textit{quantum tunneling}, where barriers in the optimization landscape are surpassed, instead of walked or sampled over.

The D-wave 2000Q QPU is composed of 2000 qubits and 5600 couplers,  with 128000 Josephson junctions. As the QPU may not have full connectivity, as it uses a \textit{chimera} architecture, so a \textit{minor embedding} is created to model the fully connected graph on the chip. For our purposes we used D-wave's software tools to automatically create a minor embedding on the QPU for our problem to be sampled \cite{ocean}.

Before approaching the problem of utilizing the quantum-enhanced selection operator, it is natural to question how well a particular solver may find energy minima for the formulation of the binary quadratic model. In order to ascertain solution quality, we randomly initialized populations of solutions according to step 1 in Algorithm 1 in Fig. 2, with which we constructed $\mathbf{Q}$ matrices according to equations 1. and 2. We then ran trials for each solver type, with the set of solvers consisting of:

\begin{itemize}
    \item D-Wave 2000Q - D-Wave Sampler (DwS)
    \item D-Wave 2000Q - D-Wave Clique Sampler (DwCS)
    \item Leap Hybrid Sampler (LHS)
    \item Simulated Annealing (SA)
    \item Steepest Descent (SD)
\end{itemize} 

For the quantum samplers, D-Wave provides tools to embedding the QUBO on the chip. In the case of the D-Wave Sampler, a minor embedding using the embedding composite tool was constructed for this purpose to map the problem onto the QPU. For the D-Wave Clique Sampler, the tool attempts to find clique embedding on the chip of equal chain length. An important parameter, \textit{chain strength}, was set as the maximum absolute value of the linear terms of the initialized binary quadratic model for both quantum samplers and embedding tools \cite{ocean}.

D-Wave also provides access to a hybrid sampler, which leverages both classical and quantum calls within it's subroutine. This technology is proprietary to D-Wave, and therefore we treat this sampler as a black box, and assume that some component of the subroutine leverages calls to a quantum processing unit. For the classical solvers, simulated annealing and steepest descent are relatively straight forward in their implementations and may be reviewed per D-Wave's documentation \cite{ocean}.

In our tests we examined varying the values of the parameters $\alpha$ and $\beta$ to see if there was any change in the  solution quality according to the distributions of minimum energies found by each sampler. Generally, we found that the Hybrid sampler achieved best performance, and the fully quantum samplers were less performant (Fig 3., Fig 4.), when run over the same set of input QUBOs. This could be attributed to the fully connected nature of the input problem, where the embedding found of the decomposition of the fully connected graph for the quantum samplers may not be optimal with respect to the QPU architecture. Since the Hybrid Sampler achieved the best results over different values of $\alpha$ and $\beta$ (Fig. 3, 4), we used this sampler for our experiments over the OneMax function and functions from IOHProfiler.

\subsection{OneMax}
In our initial experiments we benchmarked our quantum-enhanced genetic algorithm against genetic algorithms with $(\mu + \lambda)$, $(\mu, \lambda)$ over the OneMax function. 

Following \cite{SE91},the members of our population are vectors of length $\nu$, $\mathbf{a} = [x_0, x_1,,, x_\nu]$, where $x \in [0,1]$. We wish to find the bit-string $\mathbf{a}$ which maximizes the function:

\begin{equation}
    f(\mathbf{a}) = \sum_{i=1}^\nu x_i
\end{equation}
The optimum $\mathbf{a'}$ of which is essentially a vector of ones, $\mathbf{a'} = [1,1,...1]$ where $|\mathbf{a'}|$ = $\nu$.

%\subsection{Evolution Strategies and the Ackley Function}

%\begin{figure}
%  \includegraphics[width=\linewidth]{./figures/es_best_fitness2.png}
%  \caption{Average fitness for quantum-enhanced vs. classical operators for Evolution Strategies over Ackley objective function}
%  \label{fig:es_bf}
%\end{figure}

%\begin{figure}
%  \includegraphics[width=\linewidth]{./figures/es_gd.png}
%  \caption{Average genotype diversity for quantum-enhanced vs. classical operators for Evolution Strategies over Ackley objective function}
%  \label{fig:es_bf}
%\end{figure}

\subsection{IOHExperimenter }
For our experiments, we used the IOHProfiler software library \cite{IOHprofiler} for black-box optimization. IOHProfiler contains a suite of Pseudo-Boolean functions with which to benchmark optimization algorithms. The functions selected for our benchmarking include function IDs (fids) 4-18 (Table 1.). The functions 4-17 are variants of OneMax and LeadingOnes, and are \textit{W-model} transformed, using \textit{dummy variables} (\textit{DV}), \textit{neutrality} (\textit{Neu}), \textit{epistasis} (\textit{Eps}), and \textit{ fitness perturbation} (\textit{FP}).
\begin{table}[h]
\begin{tabular}{|l|l|l|l|l|l|l}
\hline
\multicolumn{6}{|c|}{Table of \textit{W-model} transformed objective functions } \\
\hline
FID & function & \textit{ DV} & \textit{ Neu} & \textit{ Eps} & \textit{ FP }\\
\hline
4 & OneMax &$n/2$ & 1 & 1& $id$ \\
5 & OneMax &$0.9n$ & 1 & 1 & $id$\\
6 & OneMax &$n$ & 3 & 1 & $id$\\
7 & OneMax &$n$ & 4 & 1 & $id$\\
8 & OneMax &$n$ & 1 & 1 & $r_1$\\
9 & OneMax &$n$ & 1 & 1 & $r_2$\\
10 & OneMax &$n$ & 1 & 1 & $r_3$\\
11 & LeadingOnes &$n/2$ & 1 & 1& $id$ \\
12 & LeadingOnes  &$0.9n$ & 1 & 1 & $id$\\
13 & LeadingOnes  &$n$ & 3 & 1 & $id$\\
14 & LeadingOnes  &$n$ & 4 & 1 & $id$\\
15 & LeadingOnes  &$n$ & 1 & 1 & $r_1$\\
16 & LeadingOnes  &$n$ & 1 & 1 & $r_2$\\
17 & LeadingOnes  &$n$ & 1 & 1 & $r_3$\\

\hline

\hline
\end{tabular}
\caption{Function transformations from IOHProfiler library \cite{IOHprofiler}, with ruggedness functions $r_1$-$r_3$ mapping various levels of fitness perturbation.}

\end{table}

Function 18 from IOHProfiler is an instance the Low AutoCorrelation Binary Sequence problem, where the fitness is determined by the reciprocal over the sequence's auto-correlation
%give the equation here
$ x\mapsto\frac{n^2}{2\sum_{k=1}^{n-1}\left(\sum_{i=1}^{n-k}s_is_{i+k}\right)^2},\text{ where } s_i=2x_i-1$ \cite{IOHprofiler}.

\subsection{Experiment Parameters, Performance Metrics and Quality Diversity }

In order to compare selection operators as part of a larger heuristic, we set some parameters within the genetic algorithm to be static across all experiments. For our choice of mutation rate, we used a rate of $m= 0.02$. For our  chromosome size, $\nu$, we set $\nu=50$ For the elitist versions of the classical and quantum-enhanced algorithms, we chose to track the best solution found so far, and recombined with the parent pool chosen by the selection operator to create offspring. In the non-elitist versions, we only used the parent pool and recombined with members of the population drawn with uniform probability. For our population size, we chose a size $n$ = 50 for all experiments, and examined the change in the size of the selected parent pool, $\mu$ in relation to the number of offspring, $\lambda$. For the ($\mu$, $\lambda$) operator, we set the number of children per parent to the value of $\lambda$/$\mu$. For settings of $\alpha$ and $\beta$, for $QE-(\mu , \lambda)$ we set $\alpha =100$ and $\beta=100$ and for $QE-(\mu + \lambda)$ we set $\alpha =1000$ and $\beta=10$.

In our experiments, we examined the expected run-time, which we define as the average number of generations to the target solution per run, with a total of 20 runs for each experiment. We set a budget of 50 generations for each experiment. We also took into account best fitness values found at each generation, which we averaged over all runs.

 %By leveraging the quantum-enhaced selection operator, we are interested in not only obtaining a high degree of selective pressure, but also an increase in the genotype diversity at each generation. This could allow to find a unique balance between volume and velocity in our evolutionary search.% 
We measured the quality at each generation by taking the average of the pairwise hamming distances of all members of the population at each generation, and averaging over these in each individual run, finally taking the average for all 20 runs.  

\section{Results}
We plotted the results as distributions of energies per sampler on random initialization of populations with varying values of $\alpha$ and $\lambda$ in figures 3 and 4.  We plotted the gap observed in the change in ratio of $\mu$/$\lambda$ vs. average generations to convergence over OneMax in figure 5. We plotted the average fitness and log of average genotype diversity over the One Max Function in figures  6 and 7. 
We tabulated performance results with quantum-enhanced vs. classical algorithms in Tables 2,3 and 4. We highlighted best performing in bold, ranked in order by average fitness, average generations to convergence, and average genotype diversity. For fids 11,12,13,18 the $QE-(\mu + \lambda)$ operator outperformed other operators in terms of fitness and average generations to convergence. For fids 7, 10, 14, 15, 16, 17 the $QE-(\mu , \lambda)$ outperformed other versions with regards to fitness values.  To verify the significance of these results we ran t-tests for independence over the samples of trials vs. their classical versions, and found all within significant range ($p-value$<0.05). 

\begin{figure}
  \includegraphics[width=\linewidth]{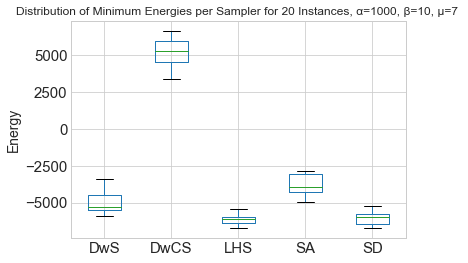}
  \caption{Distributions of energies found per sampler on randomly initialized QUBOs, $\alpha$=1000, $\beta$=10, $\mu$=7}
  \label{fig:1000alpha}
\end{figure}

\begin{figure}
  \includegraphics[width=\linewidth]{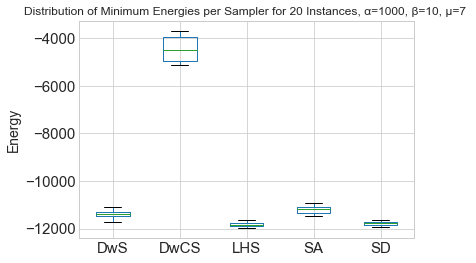}
  \caption{Distributions of energies found per sampler on randomly initialized QUBOs, $\alpha$=10, $\beta$=1000, $\mu$=7}
  \label{fig:1000beta}
\end{figure}

\begin{figure}
  \includegraphics[width=\linewidth]{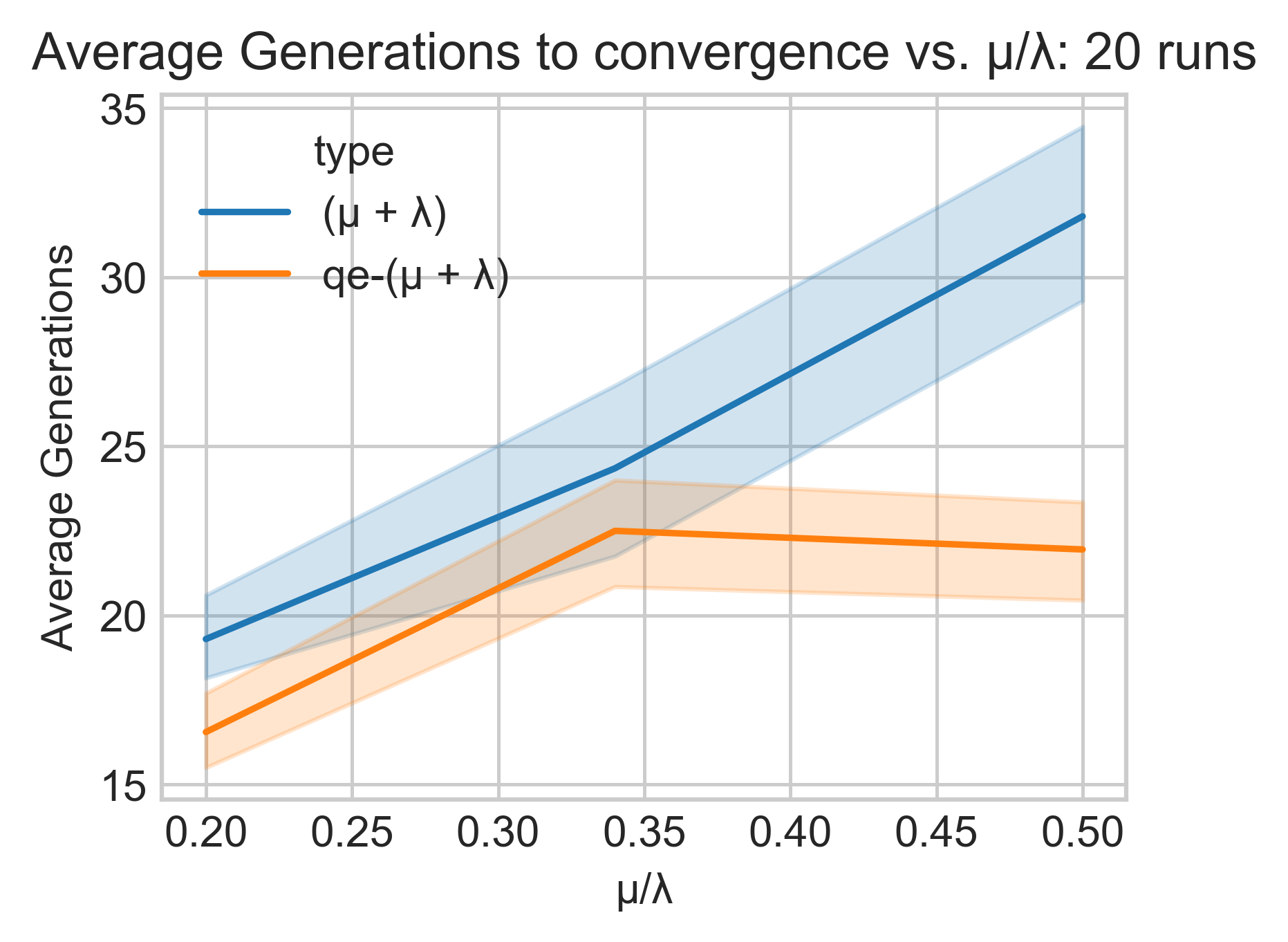}
  \caption{Change in average generations to reach global optimum for OneMax Function vs. ratio of $\mu/\lambda$ }
  \label{fig:mu_lambda}
\end{figure}
\begin{figure}
  \includegraphics[width=\linewidth]{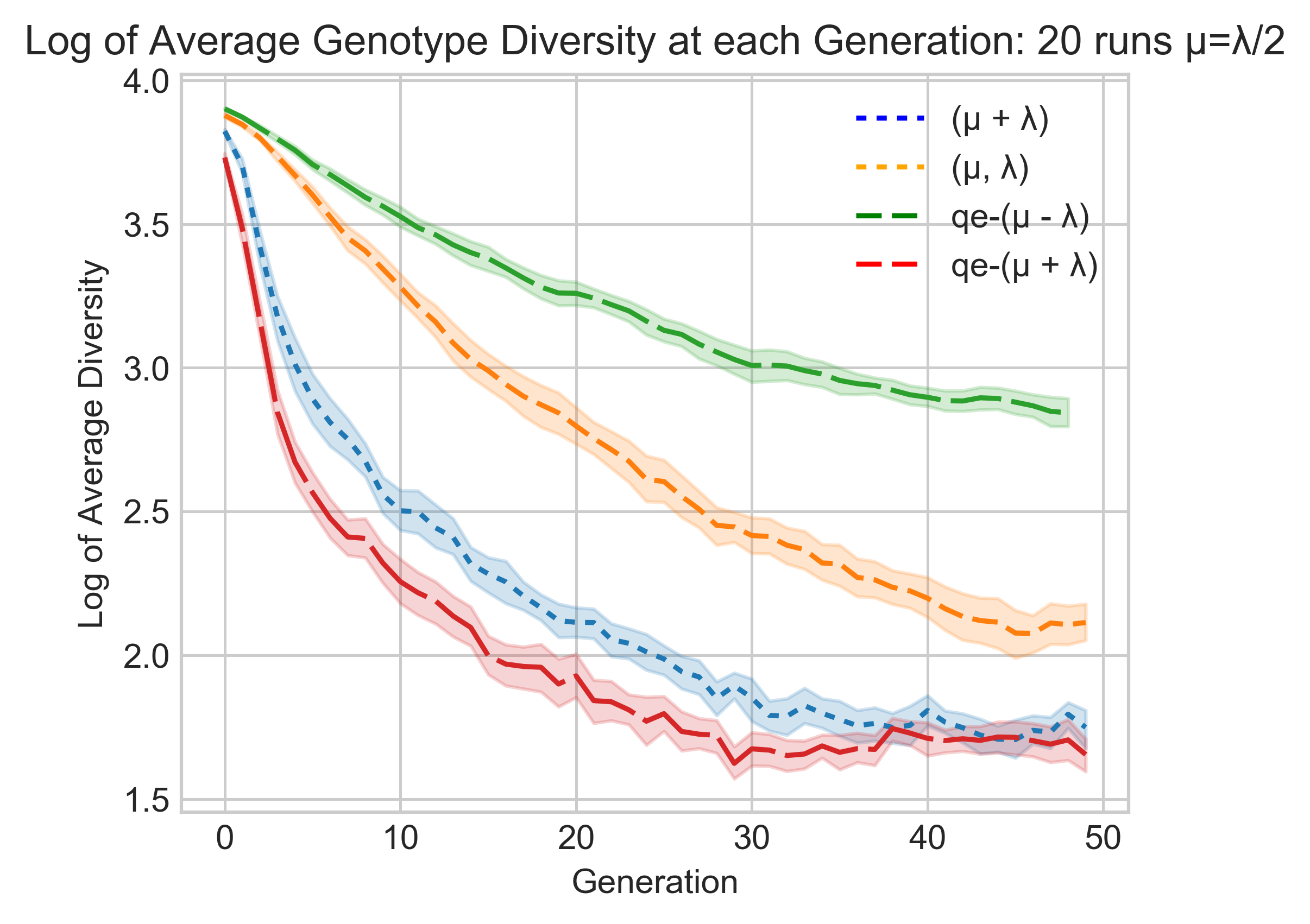}
  \caption{Average log genotype diversity for quantum-enhanced vs. classical operators over OneMax objective function}
  \label{fig:log_gd}
\end{figure}

\begin{figure}
  \includegraphics[width=\linewidth]{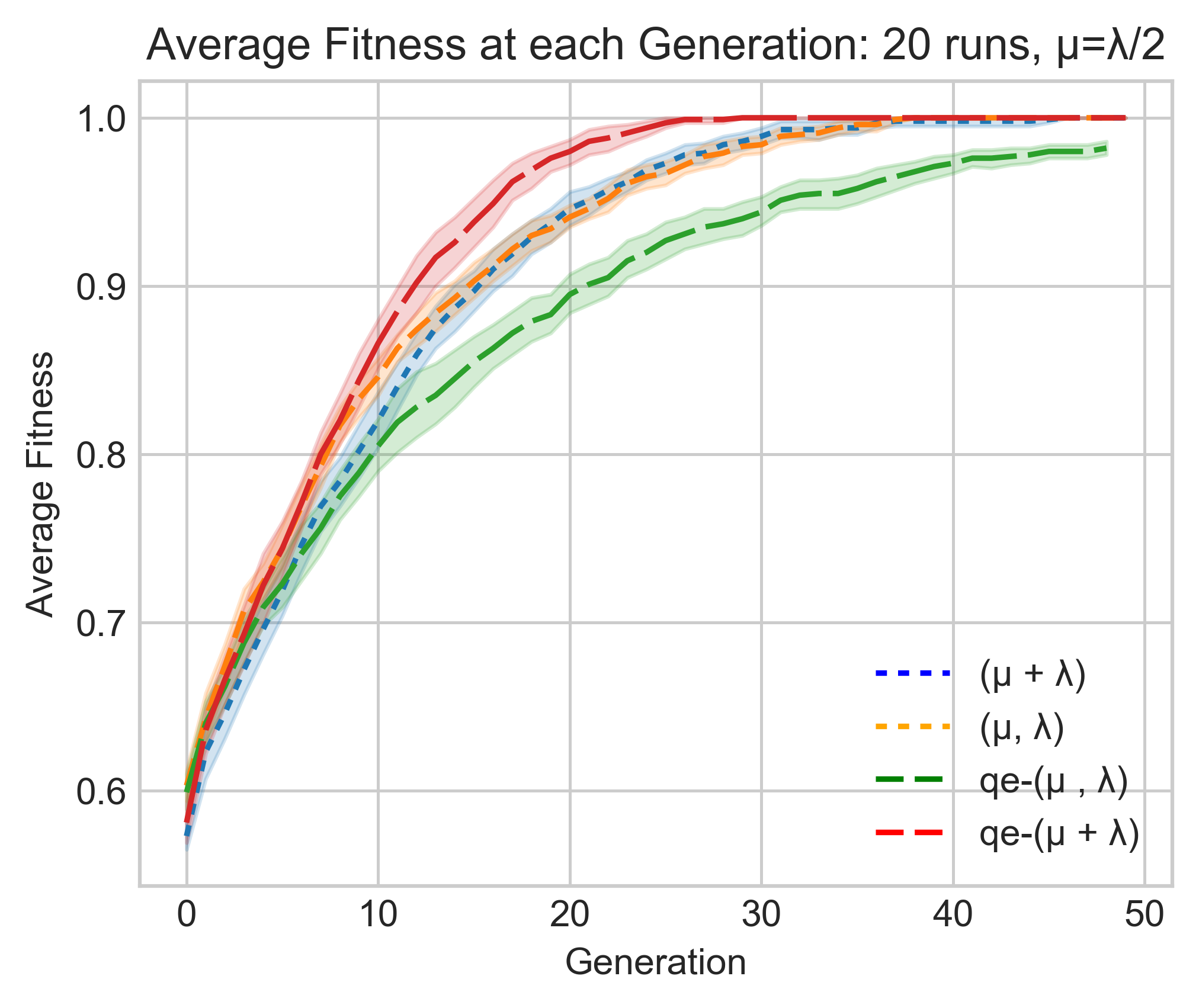}
  \caption{Average best fitness for quantum-enhanced vs. classical operators over OneMax objective function}
  \label{fig:log_best_fitness}
\end{figure}

\section{Discussion}
In examining the performance of the quantum enhanced algorithms, we plotted the average generations to convergence and genotype diversity over the OneMax function as shown in Fig. 6 and Fig. 7.  In comparing the quantum-enhanced methods to their classical counterparts we see that the quantum-enhanced  $QE-(\mu + \lambda)$ achieve high velocity towards convergence in Fig. 7., indicating higher selective pressure with lower average generations to convergence. Taking a closer look at the difference between average generations for $QE-(\mu + \lambda)$ and it's classical counterpart,  in Fig. 5, we notice a gap in expected run-time in generations as the ratio of $\mu/\lambda$ approaches 0.5. This could be attributed to higher weightings of $\alpha =1000$ vs. $\beta=10$, where preference is given to fitness within the QUBO of the quantum-enhanced selection mechanism.
We also see that there is a trade-off when comparing the average log genotype diversity of populations as indicated in Fig. 6, where $QE-(\mu , \lambda)$ shows on average a higher log diversity to it's classical counterpart. Surprisingly, when testing over the IOHProfiler suite, we notice the average genotype diversity being lower for $QE-(\mu , \lambda)$ than $(\mu , \lambda)$. This could also be due the the weighting of the $\alpha =100$ and $\beta =100$ terms, and the trade off in optimizing for fitness and lower expected run-time vs. diversity. We also see this trade off when applied to fids 4-18 in tables 2-4, where the elitist versions achieve higher fitness on functions with dummy variables and neutrality, and non-elitist versions performing well on functions with perturbations of the ruggedness function. This indicates that diversity may help to overcome these perturbations, which may lead searches with higher velocity to converge into local minima, and achieving a proper balance in weightings for terms in the QUBO when optimizing for this criterion is crucial. 

In the trade-off of exploration vs. exploitation, for elitist vs. non-elitist algorithms, we note the main difference in recombining the best solution found so far with the selected population in the elitist cases. In the cases where we tuned $\alpha =1000$, we noticed a decrease in diversity for an increase in velocity. This leads us to believe that there may be some potential in future work towards characterizing the objective function, and incorporating a switching mechanism within the optimization routine based upon whether exploitation or exploration may be more or less advantageous.  

While we used static mutation rates of 0.2, in our experiments we noticed the interplay between selection, crossover, and mutation. As we wanted to only examine the effects of the selection operator, we kept the other operators static, but we believe that future work could also examine adaptive population sizes, as well as the effect of adaptive mutation rates, which could help to reduce the diminished genotype diversity towards the end of the optimization regime as observed in Fig. 6.

%Present and discuss results

\section{Conclusion}
%Discuss conclusion and future work
We conclude that our quantum-enhanced selection operator shows some advantages in velocity and exploration within the population selection mechanism, although there is also a trade off in the latency for compute times on current NISQ chips. Future work could extend this method to Evolution Strategies to search over continuous function domains, as well as potential applications such as hyper-parameter optimization and neural architecture search for machine learning. Future work could also incorporate  streams of research previously identified in the related works section, such as Grover's search for global optimization. Finally, future work could examine quantum-enhanced surrogate modeling for both single and multi-objective optimization.

\section{Acknowledgments}
The authors declare no conflicts of interest.

\makenomenclature
\nomenclature{\textbf{$QE-(\mu + \lambda)$}}{Quantum-enhanced selection operator with elitism and recombination with best solution}
\nomenclature{\textbf{$QE-(\mu, \lambda)$}}{Quantum-enhanced selection operator without elitism, and only recombining members of selected population without tracking best solution}
%\nomenclature{\textbf{$(\mu + \lambda)$}}{Selection operator using ranked fitness values, with elitism and recombination with best solution}
%\nomenclature{\textbf{$(\mu , \lambda)$}}{Selection operator using ranked fitness values,without elitism, and only recombining members of selected population without tracking best solution}
%\nomenclature{\textbf{$TS$}}{Selection operator using tournament selection,without elitism, and only recombining members of selected population without tracking best solution}
\nomenclature{\textbf{$GD$}}{Genotype diversity, the average amount of diversity measured between genotypes in a population.}
\nomenclature{\textbf{$G$}}{Average generations to convergence or stopping criterion, also denoted as expected run-time.}

\printnomenclature

\begin{footnotesize}
\begin{table*}[h]
\begin{tabular}{|l|l|l|l|l|}
\hline
\multicolumn{5}{|c|}{IOH Function results- ( $n=50$,  $\nu=50$, 20 trials)} \\
\hline
 
IOH FID & $QE-(\mu + \lambda)$ Mean Fitness  & $(\mu + \lambda)$ Mean Fitness &$QE-(\mu - \lambda)$ Mean Fitness &$(\mu - \lambda)$ Mean Fitness\\
\hline
4 & 25 (+/- 0.0) & \textbf{25 (+/- 0)} & 25 (+/- 0) &25 (+/- 0)  \\
5 & 44.95 (+/- 0.21)& \textbf{45.0 (+/- 0.0)} & 45 ( +/- 0.)& 45.0 (+/- 0.)  \\
6 & 16 (+/- 0.0) & 15.95 (+/- 0.2) & 16 (+/- 0.0) & \textbf{16 (+/- 0.0)}  \\
7 & 43.45 (+/- 1.01) & 43.05 (+/- 1.74) & \textbf{46.65 (+/- 1.31)} & 43.45 (+/-1.65)  \\
8 & 25.25 (+/- 0.43) &  24.85 (+/- 0.57) & 25.4 (+/- 0.0)& \textbf{25.95 (+/- 0.0)} \\
9 & 49.1 (+/- 0.83) & 48.9 (+/- 0.7) & 49.6 (+/- 0.21) & \textbf{49.9 (+/- 0.3)}  \\
10 & 34.5 (+/- 4.15 &  34.5 (+/- 3.1) & \textbf{48.0 (+/- 2.0)} & 45.5 (+/- 2.29) \\
11 & \textbf{25 (+/- 0.0)}  & 25 (+/- 0.0) & 24.3 (+/- 1.3) & 23.9 (+/- 1.69)  \\
12 & \textbf{42.1 (+/- 2.9)} & 40.0 (+/- 5.2) & 25.45 (+/- 3.2) & 23.5 (+/- 3.45) \\
13 & \textbf{16 (+/ 0.0)}  & 16 (+/ 0.0)& 16 (+/ 0.0) & 16 (+/ 0.0)  \\
14 & 10.1 (+/- 4.7)&  10.25 (+/- 4.14) & \textbf{12.35 (+/- 5.8)}& 9.65 (+/- 3.2) \\
15 & 11.25 (+/- 4.3) & 11.8 (+/- 5.3) & \textbf{13.3 +/-(1.9)}  & 12.15 (+/- 2.15) \\
16 & 17.55 (+/-  7.14) & 17.4 (+/- 6.0) & \textbf{24.95 (+/- 4.2)} & 20.85 (+/- 5.1) \\
17 & 9.25 (+/- 3.34) & 9.5 (+/- 4.15)& \textbf{16.9 (+/- 5.1)} & 10.75 (+/- 4.26) \\
18 & \textbf{4.01 (+/- 0.35)} & 3.81 (+/- 0.31) & 2.79 (+/- 0.23)& 3.09 (+/- 0.32) \\

\hline
\end{tabular}
\caption{Average Fitness values for 20 trials.}
\end{table*}

\begin{table*}[h]
\begin{tabular}{|l|l|l|l|l|}
\hline
\multicolumn{5}{|c|}{IOH Function results- ( $n=50$,  $\nu=50$, 20 trials)} \\
\hline
 
IOH FID &$QE-(\mu + \lambda)$ Average G  & $(\mu + \lambda)$ Average G  &$QE-(\mu - \lambda)$ Average G &$(\mu - \lambda)$ Average G\\
\hline
4 & 10.2 (+/- 1.9) & \textbf{9.8 (+/- 2.1)} & 11.8 (+/- 1.5) & 11.85 (+/- 1.95) \\
5 & 23.2 (+/- 6.6) & \textbf{21.3 (+/- 2.7)} & 30.15 (+/- 4.4) & 26.55 (+/- 4.9) \\
6 & 12.25 (+/- 7.1) & 21.3 (+/- 15.0) & 9.7 (+/- 2.3) & \textbf{8.85 (+/- 2.55)} \\
7 & 50 (+/- 0.0) &50 (+/- 0.0) & \textbf{50 (+/- 0.0)} &  50 (+/- 0.0) \\
8 & 46 (+/- 7.45) & 49.55 (+/- 1.35) & 44.75 (+/- 8.04) &  \textbf{35.4 (+/- 5.64)} \\
9 & 44.2 (+/- 9.5) & 48.35 ( +/- 5.1) & 44.1 (+/- 6.26) &  \textbf{38.55 (+/- 7.2)} \\
10 & 50 (+/- 0.0)  & 50 (+/- 0.0) & \textbf{50 (+/- 0.0)} &50 (+/- 0.0) \\
11 & \textbf{17.85 (+/- 4.8)} & 20.75 (+/- 5.7)& 30.7 (+/- 9.3) & 43.55 (+/- 8.17) \\
12 & \textbf{48.15 (+/- 3.7)} & 45.5 (+/- 5.2) & 50 (+/- 0.0) & 50 (+/- 0.0) \\
13 & \textbf{9.85 (+/- 4.7)} & 14.3 (+/- 10.6)& 21.65 (+/- 6.5) & 22.5 (+/- 9.48)  \\
14 & 50 (+/- 0.0)  & 50 (+/- 0.0) & \textbf{50 (+/- 0.0)} &50 (+/- 0.0) \\
15 & 50 (+/- 0.0) &50 (+/- 0.0) & \textbf{50 (+/- 0.0)} &50 (+/- 0.0) \\
16 & 50 (+/- 0.0) & 50 (+/- 0.0)& \textbf{50 (+/- 0.0)} &50 (+/- 0.0) \\
17 & 50 (+/- 0.0) & 50 (+/- 0.0) & \textbf{50 (+/- 0.0)} &50 (+/- 0.0)  \\
18 & \textbf{50 (+/- 0.0)} &50 (+/- 0.0) & 50 (+/- 0.0) &50 (+/- 0.0) \\

 \hline
\end{tabular}
\caption{Average generations to convergence values for 20 trials.}
\end{table*}

\begin{table*}[h]
\begin{tabular}{|l|l|l|l|l|}
\hline
\multicolumn{5}{|c|}{IOH Function results- ( $n=50$,  $\nu =50$, 20 trials)} \\
\hline
 
IOH FID &$QE-(\mu + \lambda)$ Average GD  & $(\mu + \lambda)$ Average GD  & $QE-(\mu - \lambda)$ Average GD &$(\mu - \lambda)$ Average GD \\
\hline
4 & 0.45 (+/- 0.02) & \textbf{0.68 (+/- 0.02)} & 0.62 (+/- 0.0) &0.68 (+/- 0.01)  \\
5 & 0.4 (+/- 0.01) & \textbf{0.68 (+/- 0.02)} & 0.57 (+/- 0.01)& 0.69 (+/- 0.02)  \\
6 & 0.45 (+/- 0.04) & 0.68 (+/- 0.02) & 0.63 (+/- 0.01) & \textbf{0.69 (+/- 0.02)} \\
7 & 0.35 (+/- 0.01) & 0.67 (+/- 0.02) & \textbf{0.56 (+/- 0.01)} & 0.68 (+/- 0.02) \\
8 & 0.37 (+/- 0.01)& 0.67 (+/- 0.01) & 0.57 (+/- 0.01) &  \textbf{0.67 (+/- 0.01)}  \\
9 & 0.36 (+/- 0.02)& 0.68 (+/- 0.01) & 0.55 (+/- 0.01)& \textbf{0.68 (+/- 0.02)}  \\
10 & 0.36 (+/- 0.02) & 0.68 (+/- 0.02)& \textbf{0.59 (+/- 0.0)} &0.68 (+/- 0.02)  \\
11 & \textbf{0.45 (+/- 0.02)}& 0.68 (+/- 0.01) &0.58 (+/- 0.01) &0.69 (+/- 0.02)  \\
12 & \textbf{0.40 (+/- 0.01)} &0.68 (+/- 0.02) &0.59 (+/- 0.0) &0.68 (+/- 0.02)  \\
13 & \textbf{0.48 (+/- 0.04)} &0.69 (+/- 0.01) &0.61 (+/- 0.01) &0.68 (+/- 0.02)  \\
14 & 0.35 (+/- 0.01) & 0.68 (+/- 0.02) & \textbf{0.61 (+/- 0.01)}&0.68 (+/- 0.01)  \\
15 & 0.36 (+/- 0.01)& 0.68 (+/- 0.02)& \textbf{0.59 (+/- 0.0)} & 0.68 (+/- 0.02) \\
16 & 0.36 (+/- 0.01) & 0.69 (+/- 0.02 & \textbf{0.59 (+/- 0.01)} & 0.67 (+/- 0.02) \\
17 & 0.34 (+/- 0.01) & 0.68 (+/- 0.02)& \textbf{0.61 (+/- 0.01)}& 0.68 (+/- 0.02)  \\
18 &\textbf{0.34 (+/- 0.01)} & 0.69 (+/- 0.01) &0.64 (+/- 0.0) & 0.68 (+/- 0.02) \\

 \hline
\end{tabular}
\caption{Average genotype diversity to convergence values for 20 trials.}

\end{table*}

%\begin{table*}[h]
%\begin{tabular}{|l|l|l|l|l|l|}
%\hline
%\multicolumn{6}{|c|}{Table of objective \textit{W-model} %transformed objective functions from IOHProfiler library} \\
%\hline
%FID & function & Dummy variables & Neutrality & Epistatis & Fitness perturbation \\
%4 & OneMax &$n/2$ & 1 & 1& $id$ \\
%5 & OneMax &$0.9n$ & 1 & 1 & $id$\\
%6 & OneMax &$n$ & 3 & 1 & $id$\\
%7 & OneMax &$n$ & 4 & 1 & $id$\\
%8 & OneMax &$n$ & 1 & 1 & $r_1$\\
%9 & OneMax &$n$ & 1 & 1 & $r_2$\\
%10 & OneMax &$n$ & 1 & 1 & $r_3$\\
%11 & LeadingOnes &$n/2$ & 1 & 1& $id$ \\
%12 & LeadingOnes  &$0.9n$ & 1 & 1 & $id$\\
%13 & LeadingOnes  &$n$ & 3 & 1 & $id$\\
%14 & LeadingOnes  &$n$ & 4 & 1 & $id$\\
%15 & LeadingOnes  &$n$ & 1 & 1 & $r_1$\\
%16 & LeadingOnes  &$n$ & 1 & 1 & $r_2$\\
%17 & LeadingOnes  &$n$ & 1 & 1 & $r_3$\\

%\hline

%\hline
%\end{tabular}
%\caption{Function transformations from IOHProfiler library.}

%\end{table*}

\end{footnotesize}

\bibliographystyle{ACM-Reference-Format}
\bibliography{biblio}

%\appendix



\end{document}